# Using Word Embedding and Convolution Neural Network for Bug Triaging by Considering Design Flaws

Reza Sepahvand[1], Reza Akbari[1], Behnaz Jamasb[1], Sattar Hashemi[2], Omid Boushehrian[1]
[1]Sotware Engineering Lab, Department of Computer Engineering and IT, Shiraz University of Technology
[2]Department of Computer Science, Engineering, and IT, Shiraz University

___

**Abstract**: Resolving bugs in the maintenance phase of software is a complicated task. Bug assignment is one of the main tasks for resolving bugs. Some Bugs cannot be fixed properly without making design decisions and have to be assigned to designers, rather than programmers, to avoid emerging bad smells that may cause subsequent bug reports. Hence, it is important to refer some bugs to the designer to check the possible design flaws. Based on our best knowledge, there are a few works that have considered reffering bugs to designers. Hence, this issue is considered in this work. In this paper, a dataset is created, and a CNN-based model is proposed to predict the need for assigning a bug to a designer by learning the peculiarities of bug reports effective in creating bad smells in the code. The features of each bug are extracted from CNN based on its textual features, such as a summary and description. The number of bad samples added to it in the fixing process using the PMD tool determines the bug tag. The summary and description of the new bug are given to the model and the model predicts the need to refer to the designer. The accuracy of 75% (or more) was achieved for datasets with a sufficient number of samples for deep learning-based model training. A model is proposed to predict bug referrals to the designer. The efficiency of the model in predicting referrals to the designer at the time of receiving the bug report was demonstrated by testing the model on 10 projects.

**Keywords**: software maintenance, bug triage, deep learning, convolution neural network, word embedding, design flaw prediction, code smell

___

## 1. Introduction

Evidence from various studies suggests that maintenance activities are among the most expensive activities in the software development life cycle [1]-[2]. Palomba et al. [3] reported that maintenance costs range from 2 to 100 times the cost of software development. Providing a way to automate the bug process is very attractive due to its impact on cost reduction. Bugs reports received from Bug Tracking Systems (BTS), must be triaged. Due to a large number of bugs received, manual triaging is very time-consuming.

Extensive research has been done to automate the bug-triaging process. One of the main tasks in bug triaging is the suggestion of a good programmer to fix the bug [4]-[7]. All previous research on the bug management system has led to the help of a triager to assign the bug to the right programmer. However, it seems that the source of many software bugs is design flaws. The design flaws as a cause of software bugs have been investigated by researchers and their correctness has been demonstrated. Not correcting the design lead to the error-prone code. In this work, we are looking for a solution to decide if we need to change the design while fixing the software bug or not. If design changes are needed instead of assigning a bug to the programmer, we first refer it to the designer to review the design flaws.

In this study, the need to change the software design in the bug fixing process by receiving a bug report is examined. To achieve this, we must first define the bad design and its criteria. For this purpose, the criteria of a code with bad design are determined. This type of code should be referred to a designer to examine the design and probably correct it. The version for fixing each bug is reviewed according to the criteria and the need to examine the design is investigated. This review and compliance with poor design criteria should be done before the bug is referred. A model should be provided that receives the new bug and poor design criteria as the input. The model predicts code changes and identifies whether or not to examine the design.

Bad smell is one of the hallmarks of a bad design. To detect bad smells, some signs are defined and the code that has these signs may have a design flaw and this code should be examined in terms of design. In this work, we will deal with bad smells as signs of bad design in the software code, which forms the basis of the current research. The article on code smells was provided by Fowler [8]. Bad code or bad smells are a sign of poor design and implementation, which has a bad effect on the work of developers [9]-[10]. They remain in the software for a long time and have side effects such as increasing change times, making fault-prone code [11]-[12], reducing comprehension [13], and reducing the maintenance capability [14]-[16], [8].

Bad smells may not directly lead to software failures. However, they may indirectly lead software to fail [17]. They make software changes harder, which in turn can lead to bugs. There is a lot of research that demonstrates the correlation between the code smell and defect-prone codes. Various works have proposed that using the code smells in the input of the bug prediction models increases their efficiency [18]. In [19], several projects have empirically concluded that design flaw correlates with software defect. Increasing the number of design flaws increases the likelihood of bugs occurring. Code artifacts that suffer from bad smells due to maintenance and evolution have different characteristics than clean samples. This finding is consistent with the results of the history of the evolution of bad smells in [20]-[22].

However, there are cases, especially in blob and complex class, where bad smells appear on the file after several changes. In such cases, files having bad smells reveal a certain tendency toward quality metrics that are completely different from clean files. These results encourage the development of advisors who can warn software developers in a timely manner. The increased worrying parameters due to the code changes lead to bad smells in the code[12].

According to the mentioned results, it seems that the design modification in the software maintenance stage is necessary. However, based on our best knowledge, there are a few works that have been done to help the triager to assign a bug to the designer. Also, there is a high correlation between the code smells and the software bug, and codes with bad smells have features that can be used to predict a bad smell. Further, code changes in the bug-fixing software version are predictable when the bug is received. Finally, with software changes, bug-prone modules are predictable.

This motivates us to provide a way to predict the number of bad smells by receiving a bug report. This is performed before making any changes in the source code and without seeing the software code. The goal is to suggest a model that predicts the bad smell appearance in source code in the bug fixing process. In fact, by receiving the bug report, and before assigning it to the programmer and making a change in the source code, we predict what bad smells it will have in the bug fixing process. In this way, we refer the bug to the designer to fix the design defects. This model uses information about previously fixed bugs such as the software bug reports' information and bad smells that exist in the software source before and after fixing the bug in the training phase. The applied model uses only the reported bug information such as summary and description for predicting bad smells and there is no need for software source code. Based on our best knowledge, predicting the extent of bad smells in the bug fixing process is proposed for the first time. This is done in the first stage of bug triaging by receiving a bug report and before bug assignment, by seeing the bug report and without the need for software source code. By predicting the appearance of the bad smell in the bug fixing process, if the bad smell appearance is higher than a specific threshold, refer the work to a designer instead of the programmer to fix design flaws to decrease future bugs' occurrence.

The most important contributions of this research are as follows:

- Providing a solution to predict the need to change the software design in the bug fixing process by receiving a bug report to help the triager to refer the bug to the designer.
- Predicting the number of bad smells that appears in the modified software code to fix the received bug report, using bug report information such as summary and description, as soon as the report is received before the bug is installed and before changes are made to the software and without the need to have a modified source code in the bug fixing process.
- Providing a deep learning method and extracting textual features and classifying the bug into two classes: 1) the bug needs to be referred to the designer, and 2) the bug does not need to be referred to the designer.
- Preparing a dataset for the reported bugs of 10 projects and specifying its class labels with two labels: need to refer bug to the designer or no need to refer. This is done by analyzing the code smells of the source code before and after fixing the bug using the PMD tool.
- Carrying out an empirical study on 10 different projects and proving the efficiency of the proposed method to predict design flaws.

The continuation of this research is as follows: The second section deals with related work. In the third section, the overall view of the proposed method is presented. Section 4 presents data preparation. Section 5 presents the prediction model. The sixth section analyzes the results and finally Section 7 concludes this work.

## 2. Related Work

In recent years, many research works have been reported in the bug triaging domain. Habayeb et al. [23] proposed an HMM-based method for classifying bugs into short and long fixation times. In their method, the activities performed in the bug fixing process were extracted. Based on whether the bug fixing time was longer than the average fixing time or less, they were divided into two categories. An HMM-based model was created for each category, and a new bug was given to the two models to determine their class. The model was evaluated with Firefox data and its higher efficiency has been demonstrated. In our previous work [24], we proposed a method based on deep LSTM and Word embedding to improve the work done by Habayeb et al. [23]. In the proposed method, by using word embedding, the semantic connection between the terms is recognized, and by using Deep LSTM, the connection between Long term and short-term is discovered. The results of the evaluation of the algorithm on Habayeb et al. dataset showed 15% improvement.

In recent years, a lot of research has been done on fault localization to predict the part of the software code that causes the bug. The basis of most methods is to find similarities between the text of the bug report and the source of the software using IR-based methods [25]. Gharibi et al.[26] proposed a method for bug localization by combining a series of methods. Each method provides a ranked list, and the combination of these ranked lists is presented as the final ranked list. The input of the algorithm is source code, the previously fixed bugs, and the new bug. Token matching performs bug localization by finding tokens in the text of the bug report and source code. VSM using rVMS provided in Zhou et al. [27] calculates the similarity between the bug and the source file and presents its ranked list. Stack trace provides a list using the stack trace in the bug report. The semantic similarity component fills the gap between natural language and programming language due to the differences between the terms in the source code and the bug report. This component also provides a ranking list. The fixed bug reports component provides a list based on modified files in previously fixed bugs. The final list is generated by combining the lists provided by the various components. The proposed method has been evaluated using data from 3 open-source projects and its high efficiency has been demonstrated.

Rath et al. [6] proposed a method for bug localization by using the requirement graph and the graph of the modified files in the bug fixing process. The method was compared with SimiScore and CollabScore methods and showed a relative improvement. Zhou et al. [27] proposed the BugLocator method for finding files that caused bugs. In this method, they created a graph where its root is the new bug, the first level is the bug similar to the root bug, and the second level is the graph of the files modified in the process of fixing bugs of level 1. The similarity between the bugs is calculated using the modified version of VMS called rVMS. Based on this graph, a ranked list of possible files for bug locating is created. Another list is based on the similarity of the bug with the software source files. The final list is created by combining these two lists. The proposed method was tested with 3,000 open source project bugs, and its performance was improved compared to state-of-the-art methods.

A comprehensive empirical study was conducted by Tufano et al. [28]. They found that most bad smells occur when a file is created. However, there are cases, especially in the case of Blob and Complex Class, where bad smells appear on the file after several changes. In this case, the files having bad smells reveal a certain tendency toward quality metrics that are completely different from clean files. Smells are generally created by developers when upgrading to a new feature or implementing a new feature. As the smells usually appear 1 month before the deadline, a significant number of them will appear in the first year of the project. Developers who cause bad smells are the owners of the bad smelly files that create bad smells when they have a lot of workloads. More than 80% of bad smells have not disappeared during the evolution of the system. Only 9% of bad smells are removed as a result of refactoring operations. 40% are eliminated by deleting a bad smelly product, and adding a new code, eliminates 15% of bad smells.

Kessentini et al. [29] proposed a solution based on the distributed optimization problem to bad smell detection. The proposed method has 2 algorithms. The first one is an evolutionary algorithm based on genetic programming to generate rules for detecting code smells. The design parameters and some bad smelly codes are given to the algorithm as input. The genetic algorithm generates the number of rules equal to the number of bad smells that could be detected. The second algorithm that runs in parallel with the first is the genetic algorithm, which uses well-designed code samples to generate instances of the code smells (these are called detectors). The GP algorithm checks the coverage of the sample input code smells of the algorithm with the generated output rules of the algorithm. The GA algorithm evaluates the deviation of detectors generated by the algorithm with well-designed codes that are the input of the algorithm using global and local alignment techniques [30]. The two algorithms interact with each other using the second component of the target function called the intersection function. The function works in such a way that the intersection of code smells found by the two solutions is maximized. The best rules and detectors can be used to evaluate new systems, and there is no need to re-run the algorithm.

Xuan et al.[7] proposed a CNN-based approach for bug localization. The main focus of the work is to train the network with the data of one project and test it with the data of another project. This method is used in places where the data of a project is not sufficient for training. The proposed method was called TRANP-CNN and it was trained using fixed bug reports and bug files specified in the process of fixing them in the source project. They tested the model with destination project data, and the performance specifying a ranked list of 1, 5, and 10 suggested files are 29.9%, 51.7%, and 61.3%, respectively.

Xiao et al. [31] to use the semantic information contained in the bug report and source code, proposed the DeepLoc method for Bug localization. In the proposed method, the text of the bug report and the source code are converted to vector using word embedding. Using the CNN network, the features of these texts are extracted. An enhanced CNN is trained based on previous reports that have been fixed and files have been identified as bug location. This model has 10% better MAP than state-of-the-art methods for predicting bug files in AspectJ, Eclipse, JDT, SWT, and Tomcat bug reports.

Ruan et al. [32] proposed a deep learning method to recover links between commits and bug reports when not done manually. In the proposed method, using word embedding, the source code and the bug report are independently converted into two vectors, and their features are extracted in this way. A deep LSTM network is created and trained with previously fixed bugs and their corresponding commits. This model is used to recover links between the bug report and the source code corresponding to its fixation. The performance of this method compared to FRLink shows an improvement of 18.3%.

Xi et al. [33] Proposed a sequence-to-sequence model, called iTriage for developer assignment. The method was developed to simultaneously use the text of the bug report, metadata, and tossing graph. During the bug fixing process, developers communicate with each other by exchanging comments. In addition, they can assign the fixing task to another, which is called the tossing graph sequence. The reason for this could be a mistake in the initial assignment or using the experience of more experienced people to fix the bug. The iTrige method considers the connection between the fixers specified in the graph tossing. The iTriage consists of two models: a model for extracting features and a model for proposing the fixer. The first model includes an encoder for representing the text of the bug report using bidirectional RNNs and a decoder for tossing Sequence modeling using RNN. The Fixer Suggestion Model is a strap class that is the appropriate class for the developer bug to fix it. Evaluation results show that iTriage improves state-of-the-art methods.

3. **The Proposed Method**

In this section, we describe the proposed method for predicting the need to refer a new bug to the designer. Figure 1 explains the general structure of the proposed method. As shown in the figure, the method has two main phases:

1) Data preparation which contains data collection and labeling steps.
2) Proposign a prediction model and its track to predict whether the bug should be referred to the designer or not.

Based on our best knowledge, there are a few works that considered referring the bug report to the designer. Hence, we need to prepare a dataset to evaluate the proposed model. For this purpose, the required features of the model are extracted from the BTS system and the corresponding label from the bug is extracted from the Configuration Management System (CMS). The details of preparing the dataset are given in Section 4.

By preparing the dataset, we are ready to design a prediction model. Here, a CNN-based model is trained with this data, and its efficiency is proven by testing it. The extraction, training, and testing methods are discussed in detail in Section 5.

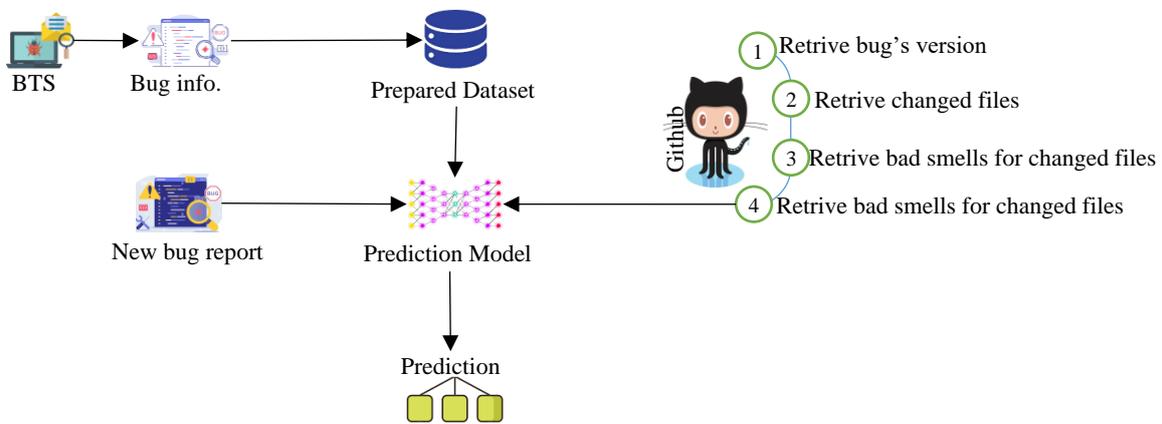

Figure 1. The overall structure of the proposed method

4. **Dataset Preparation**

The preparation of the dataset to evaluate the prediction model consists of two parts. The first part is to extract the features and the other is to specify the label for each bug. This information is extracted from previously fixed bugs in ITS and software versions related to their fixing in the GitHub system. To generate the dataset, we developed a tool. Figure 2 shows the general structure of the developed tool for dataset preparation.

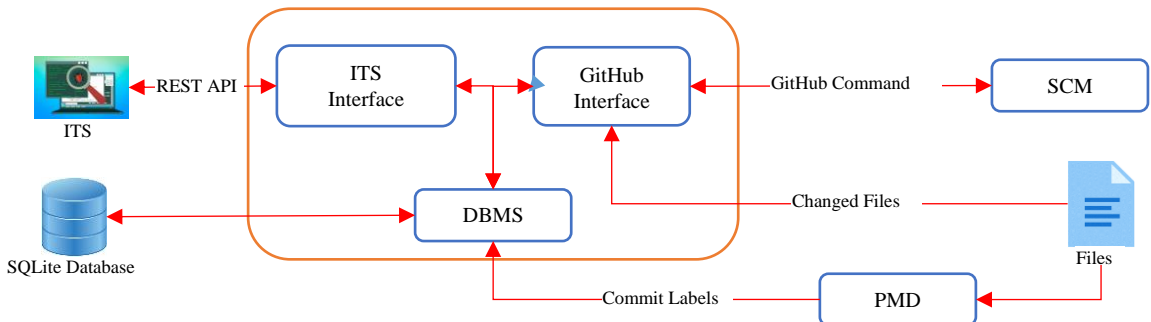

Figure 2. The developed tool for dataset preparation

**4.1 Data Collection**

In this section, we describe how to collect bug data from ITS and SCM systems and store them in a database. This information is used to extract the features and apply them to the predictor model. The ITS system is provided to manage the report and fix the bug. This system allows the user to send the report of the observed bug. The user enters information such as the date of creation of the report, the name of the reporter, the priority of the reported bug, the severity, a summary text about the bug, and a descriptive text about it. The reported bug received by the triager, is prioritized and assigned to the relevant programmer. Some information such as the date of the assignment, the assigned programmer, and the date of fixation in this process are recorded in the ITS. The programmer that the bug is assigned to may assign the bug to someone else, exchanging files or comments with others, and all of these activities and their history are recorded in the ITS system.

The information stored in this system is accessible through interfaces such as the REST API. Another system is the configuration management system (CMS) which maintains the software changes. Each version of the software is provided with a unique commit hash. In the commit provided to handle an issue, the number of this issue is recorded in the system. So, a commit hash with an issue ID are identifiers that specify the software version corresponding to fixing the bug.

SCMs such as GitHub, allow you to download different versions knowing it as commit hash. They also allow you to download modified files belonging to that version. To build a database, from the data provided by Rath et al. [6] we use and receive the required additional information from SCM and ITS and customize it according to our needs. The tables are prepared in such a way that the information about the bug is obtained from ITS and combined with the information related to the software version used to fix the bug used by SCM. With this information, more comprehensive information about the bug can be obtained and better analysis can be done.

The basic database of Rath et al. received and software in python language has been developed to manage the database. This software is used to perform various queries and apply different ACIDs to the database. In addition, the software receives the required data from ITS and stores it in the database. SCMs have a user interface that manages different versions and receives the required information. Because for each bug in all 10 projects, we need to extract the software version related to it to analyze its changes, and due to a large number of bugs and projects, it is very time-consuming to do it manually. A module has been added to the python application to automatically execute GitHub commands for each bug and extract software versions and changes made to fix the bug. Finally, in addition to storing information in the database, for each bug, the modified files in the software version provided for fixing it are stored in a folder called that bug.

As shown in Figure 2, the created database contains a set of tables. Table 1 describes the information in these tables. The most important of these tables is *Issue*. This table stores the information about the reported Issue. The information in this table is extracted from ITS. The next table is the *code_change table*. This table stores information about different versions of the software. The ID number of each version is *Commit_id*, which is the key to this table. Other information, such as files modified in the version, along with the lines of code added and subtracted in the software version, is available in this table. The next table is the *change_set_link* table. This table is created to make the connection between the two previous tables and to make it possible to collect data related to bugs. The information in this table is extracted from SCM. It has two fields, one is the *Commit_Hash* version ID number and the other is the *Issue_id* field of the bug ID, which is provided to fix that bug.

Table 1: Database information of each project

| Table name | Table Description | Field | Fields Description | Field Sample |
|---|---|---|---|---|
| code_change | The information of files that changed in software commits | Commit_Hash | The ID of the commit | edf26b4b2b755648296e1a9304859399d10b7c8c |
| | | File_path | The path of the file | src/org/apache/pig/backend/hadoop/executionengine/mapReduceLayer/partitioners/WeightedRangePartitioner.java |
| | | Sum_added_lines | Number of lines added to file in the current commit | 55 |
| | | Sum_removed_lines | Number of lines removed from file in current commit | 116 |
| change_set | Table for saving date of commits | Commit_Hash | The hash of commit | efc1198d80215a05e33b8a88a0f4612d401aae45 |
| | | Committed Date | The date of commit | 2010-07-29T21:02:29Z |
| change_set_link | Table for connecting bug and version of the software that fixed it | Issue_id | The id of the bug | PIG-733 |
| | | Commit_Hash | The commit of the software version that fixed the bug | edf26b4b2b755648296e1a9304859399d10b7c8c |
| Issues | Bug information extracted from ITS | Issue_id | The ID of Issue | PIG-733 |
| | | Issue_type | The type of Issue: Bug, New feature | Bug |
| | | Create_date | The issue creation date | 2009-03-25T19:04:22Z |
| | | Fixed_date | The issue fixed date | 2009-04-10T02:36:14Z |
| | | Summary_stemmed | The issue summary after stemming | pig 733 order sampl dump entir sampl hdf caus df file system close error larg input |
| | | Description_stemmed | The Issue description after stemming | order sampl job sampl input creat sort list sampl item urrent number item sampl 100 map task input larg result map say 50 000 sample big sort sample store df weight rang …….. |

## 4.2 Labeling

In this section, we show how to assign a class label to each bug. The class label shows whether a new bug needs to be referred to a designer to improve the design or not. To do this, the modified files in the fixing process are checked and analyzed for the presence of bad smells, and the label is determined according to the result of the analysis. Different types of code smells have been studied by presenting various signs of them, to facilitate their detection [30] and different solutions have been provided to improve their detection. The process of identifying a smelly code involves finding pieces of code that violate the properties of a structure or semantics, such as those related to coupling or complexity. In this process, the internal attributes required to define these properties are defined through software criteria, and the property is expressed through valid values for these criteria [34]. The most widely used criteria are theories defined by Chidamber and Kemerer [34]. These parameters are highly correlated with software bugs in most empirical studies. This refers to the general belief of the software development community that writing modules with large code and high complexity are a bad practice. We used bad smells to create a label for each bug. The bad smells used in this study for specifying labels for bugs are listed in Table 2.

We used the PMD tool to determine the bad smells [35]. The label of each sample is determined using the extracted bad smells. In the database, the *change_set_link* table described in Table 1, the *CommitHash* version of the software is extracted according to the bug. Information about the bad smells of each version is stored in a table. The fields in this table and their descriptions are described in Table 2. This table is extracted using the PMD tool on the modified files in the process of fixing each bug. The label for each bug is extracted using the information in this table.

Table 2: Table for storing bad-smell values for changed files in each commit

|    | Field | Description |
|----|-------|-------------|
| 1  | Commit_Hash | A commit hash identifies a commit. At the same time, it serves as a checksum to verify the integrity of the stored software object. A commit hash is 40 characters long. |
| 2  | Commit_Date | The date for the committer field |
| 3  | File_Path | It points to the location of the file. |
| 4  | Bug_ID | **The unique identifier of the reported bug** |
| 5  | AbstractClassWithoutAnyMethod | Abstract classes which do not have any methods are like simple data containers. Maybe it is better to don't make the class abstract and use a private or protected constructor. |
| 6  | CouplingBetweenObjects | The number of unique attributes, local variables, and return types within an object is counted by this rule. If the result number is higher than the specified threshold, it means the degree of coupling is high. |
| 7  | CyclomaticComplexity | This rule counts the number of decision points in a method (plus one for the method entry) to determine the complexity of a method. |
| 8  | DataClass | Data Classes have no complex functionality, they just hold data. This can demonstrate that their functionality is defined elsewhere. |
| 9  | ExcessiveClassLength | This indicates that classes have excessive responsibilities. These responsibilities can break apart into other classes or functions. |
| 10 | ExcessiveImports | In this rule, the number of unique imports is counted and if the count is above the specific threshold, a violation is reported. |
| 11 | ExcessiveMethodLength | This rule illustrates that methods have excessive responsibilities. Methods should do what their name suggests not more than it, because readers may lose their focus. |
| 12 | ExcessiveParameterList | Methods that have many parameters, especially methods that share the same datatype show the wrong situation. It can be solved by wrapping these parameters into new objects. |
| 13 | ExcessivePublicCount | This rule counts classes that have large numbers of public methods and attributes. Such classes cause combinational side effects and decrease testability. |
| 14 | GodClass | God classes are very big and have lots of functionality. This rule detects these classes so they can be separated. |
| 15 | LoosePackageCoupling | This rule shows the violation of using classes from the configured package hierarchy outside of the package hierarchy. |
| 16 | NcssCount | This rule counts the number of lines of code in a constructor, method, or class. This enumeration is done by using the NCSS (Non-Commenting Source Statements) metric. |
| 17 | NPathComplexity | The number of full paths from the beginning to the end of the block of the method is counted by NPath. The number of acyclic execution paths within a method is shown by NPath complexity. |
| 18 | SwitchDensity | A switch statement is overloaded if it has a high ratio of statements to labels. The statements can be moved to new methods or some subclasses can be created based on the switch variable. |
| 19 | TooManyFields | This rule counts classes that have Too many fields. Such classes can have fewer fields by wrapping related fields in new objects. |
| 20 | TooManyMethods | Too many methods in a class are a sign of high complexity and a way should be found to have more fine-grained objects to reduce its complexity. |

Row 1 of table *CommitHash* specifies the version number. Changed files in each version are specified with row 3 (*File_Path*). The fourth row specifies the name of the identifier of the related bug to *CommitHash* version. These files are checked for 16 types of bad smells. Lines 5 to 20 of this table are the names of the fields that indicate the existence of a specified bad smell. For example, line 14 is the *GodClass* field, and if the file with the path specified in File_Path (the third row of the table) in the *commit_*Hash version(the first line field) has God Class, this field has saved 1. Similarly, 16 bad smells are stored for each file in each commit. The fields specified in lines 7 and 17, respectively, *CyclomaticComplexity* and *NPathComplexity*, have numerical values, which are converted to 0 and 1 by comparing them with the threshold level of 40, and thus have 0 or 1 value in the calculations of all fields. The fields for bad smells are also checked in the previous version. The previous version was identified using the *change*_set table, which is described in Table 1. If a file in a version has a specific bad smell and this bad smell does not exist in the previous version - added in the current version - it will be registered as the added bad smells.

$$HasBadSmell(\text{CurCommit}) = bool\left(\sum_{n=1}^{\text{changed File in Cur Commit}} \left(\sum_{j=5}^{20}(Field(j.CurHash) - (Field(j.PrevHash))\right)\right) \quad (1)$$

The bad smells added in the current version are added together for all the modified files. If the sum of the bad smells of a version is more than zero, it is predicted that the fixing of that bug will lead to a bad smell, and as a result, a change in design is necessary and should be referred to the designer. So label 1 is assigned to the bug. Other examples are labeled 0. The formula (1) shows the class calculation for a version. $Field(j.CurHash)$ returns the value of field number j of Table 2 in version $CurHash$. For example, Field(5.00458580c62abe7b2c1bffd1bcfaf834e5722c51) return the value of *AbstractClassWithoutAnyMethod* in version 00458580c62abe7b2c1bffd1bcfaf834e5722c51. Figure 2 illustrates the pseudo-code for specifying a label for each bug.

---

**For each** of the 10 different projects:

    **For each** bug:

        a) **Extract** the relevant commit number from the table

        b) **Extract** the relevant source code from GitHub

        c) **Extract** the modified files in the mentioned commit d) extraction bad smells for the files in the current and previous versions using the PMD tool

        e) **Assign** a label for each bug according to formula 1. Analyzing bad smells and determining the design or implementation class for the sample data (each commit is assigned a label 1 if each bad smell is added to any of its modified files) if it has a bad smell and this bad Smell also exists in the previous version, it is not considered. Only bad names are considered that have been added due to changes in this version.

---

Figure 2: the pseudo-code for specifying the label for each bug

## 5. The Prediction Model

After preparing the dataset, the prediction model is designed and applied to the data. The aim is to predict whether we need need to refer a new bug to the designer or not. Figure 3 shows the main steps of the proposed prediction model. The goal of this model is to find the best mapping function that captures the description and summary text attached and returns 1 on the need to refer to the designer and 0 otherwise. Formula (2) illustrates the mapping function. This model uses the required fields from the prepared dataset. For training the model, we must first extract the features of each bug. To extract features, a feature vector is extracted using text fields such as Description and Summary. The text data must be converted to number vectors and fed to the CNN-based network.

$$f(Description + Summary) = \begin{cases} 1 & \text{if bug need reffering to designer} \\ 0 & \text{else} \end{cases} \quad (2)$$

### 5.1 Text to Vector Using Word Embedding

In extracting textual features using CNN, since there are textual values in the input, the text must be converted to a vector. Keras library allows you to use word embedding. The embedding module is a supervised method that prepares the facility to train a neural network to convert words to a specific vector for text classification. The input of this network is different texts and the class related to them, and the output of the network is a vector that matches the words. Since the embedding and Keras input is in the form of numerical vectors, a vector is first created for each document. The Doc2Indices module does this. This module creates a dictionary from the words in the input text. It then extracts its unique words from the text of each input sample. From these unique words, a vector is created, and instead of the word, its index is inserted in the resulting vector dictionary. The formula (3) shows the process of the function. The text attached to the description and summary of bug i is called $D_i$.

$$\text{Doc2Indices}(D_i) = (a_1..a_k): k: number\ of\ unique\ words\ in\ D_i. \quad (3)$$

The numeric vector corresponding to each document is given to the embedding layer. The other input of the embedding layer is the class label corresponding to each bug that is provided in the Y vector of this input. The embedding layer is a neural network itself, and the output of this layer is a matrix. The rows of this matrix are equal to the number of unique words in the document. Each row of the matrix is a vector corresponding to the mapping of the word dictionary index in the q dimension space. q is equal to the dimensions of the space to which we want to map the word. Here we consider the number of dimensions equal to q = 128. Assuming that the number of unique words is equal to the word p, then the output of the Embedding step is the $X_{p \times q}$ matrix.

$$M = \text{Embedding}\left(\text{Doc2Indices}(D_i)\right) = \begin{bmatrix} w_{1.1} & \cdots & w_{1.128} \\ \cdots & \cdots & \cdots \\ w_{p.128} & \cdots & w_{p.128} \end{bmatrix} \quad (4)$$

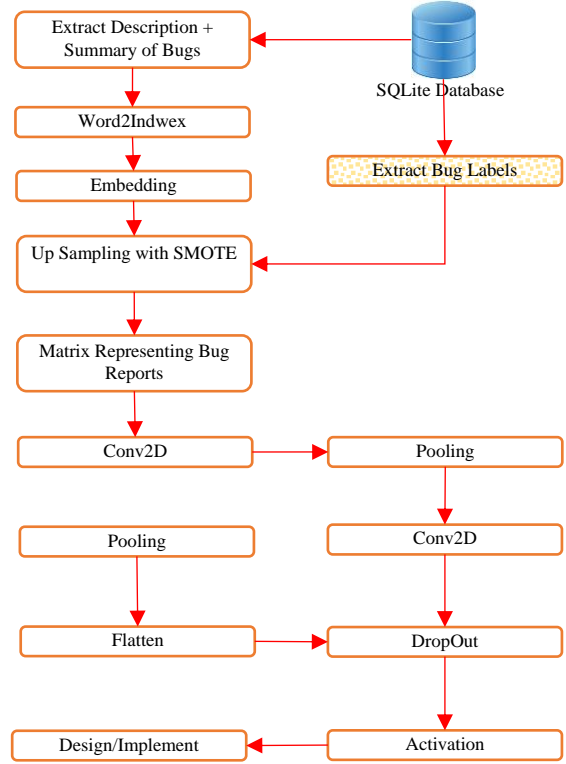

Figure 3: The proposed prediction model

### 5.2 Feature extraction using word embedding and CNN and bug classification

In ITS, we choose the features that are most relevant despite the design flaws. The text of the bug report has been used in many studies in various fields, such as predicting the time of fixing the bug, predicting the location of the bug, selecting the programmer, and proving its efficiency. So the bug report text is selected as the most important feature in bug-related studies. A CNN-based model is used to extract text features from the bug description and summary fields. Using Word2Index and word embedding, the text of the report is converted to a vector of numbers. This vector is given to a deep CNN. The network has two stages of convolution and Pooling, one phase of Flatten and Dropout to prevent overfitting, and finally an activation function for two-class classification. Formula (5) shows the process.

$$pred(M) = sig(DropOut(Flatten(Pool\left(Conv\left(Pool(Conv(M))\right)\right)))) \quad (5)$$

### 6. Performance Study

In this section, we analyze the dataset and the performance of the prediction model. Data from 10 open source projects are used to evaluate the model. We will first analyze the dataset. The effect of the changing K in K-Fold validation is evaluated, and at the end, the efficiency of the method is evaluated on 10 different datasets.

### 6.1 Dataset analysis

As described in Section 5, for the description and summary of each bug, its features are extracted using a CNN-based model. To each bug, a label is extracted. After assigning the label to each bug, which is assigned to that label 1 if referring the bug to the designer is needed and 0 otherwise. The frequency of need for referral to a designer is on average 30%. Table 3 illustrates the frequency of different classes for different projects. In this table, for each project, the total number of samples, the number of samples required to be referred to the designer with label 1, and the number of samples to be referred to the implementer with label 0 is specified. The percentage of samples that need to be referred to the designer is also specified in the last column. This percentage averages 36.7 in 10 projects.

As can be seen, the number of samples with label 1 is much lower than the number of samples with label 0 and the data is imbalanced. Since machine-learning algorithms have been developed and optimized to train with balance data, the use of imbalanced data leads to their bias towards more frequent classes and ultimately causes model overfitting. To solve this problem, the data should be balanced. Data balancing is done in two ways: Up sampling and Down Sampling. In the downsampling method, some samples with a class label with greater frequency are removed to equal the number of samples with different classes. This method leads to the loss of suitable samples for training and the reduction of training data. In the Up Sampling method, fewer class samples are created synthetically to balance the number of positive and negative samples.

Table 3: samples with label 0 and 1

| # | Project | Num. of class 1 Samples | Total sample | Class 1 Percent |
|---|---|---|---|---|
| 1 | Derby | 1101 | 4789 | **23** |
| 2 | Infinispan | 1071 | 3977 | **27** |
| 3 | Teeid | 888 | 3797 | **23** |
| 4 | Log4j2 | 342 | 1165 | **29** |
| 5 | Drools | 622 | 1121 | **55** |
| 6 | HornetQ | 382 | 1039 | **37** |
| 7 | Seam2 | 355 | 1025 | **35** |
| 8 | Railo | 217 | 696 | **31** |
| 9 | Izpack | 209 | 844 | **25** |
| 10 | Wildfly | 181 | 220 | **82** |

In this study, we used the Up Sampling method and updated the data using the Synthetic Minority Oversampling Technique (SMOTE) method proposed by Chawla et al. [36]. It is still an effective method in this regard. This method is also used in text classification [37]-[40]. This method is still very popular and its variants are offered [41]-[42]. This method is applied to data with continuous numerical values. After converting the text to a vector, it has these properties and this method can be applied to them. In this method, one of the samples is selected with less class frequency so that new samples can be made from them. For each sample, its K neighbor is selected. One of these samples is randomly selected. Multiplying a random number between 0 and 1 in this vector creates a new vector that is added to the data set as a new instance. This process continues until the data is balanced.

To evaluate the proposed model, it must be trained with part of the dataset and tested with the rest of the data. We used the K Fold validation method to evaluate. K determines the percentage of data for testing. For example, in 10 folds, we divide the data into 10 parts, and each time we select one part for testing and train the model with the remaining 9 parts. This method is repeated 10 times and the average of parameters is reported as a result of the evaluation. To choose the best K, the model must be evaluated with different K. Since it takes time to train deep learning-based models, to select the appropriate K, a project with the number of suitable samples for deep learning (models based on deep learning require a lot of data for proper training) is selected and the model with Ks rated 10, 5, 3 and 2 are evaluated. For higher K, the number of training samples is higher and the model is better trained, and on the other hand, the number of test samples is less and the stability of the results should be checked. Table 4 shows the results of the implementation of the Infinispan project. Given that there are some proper samples in this project, changing K does not have such an effect on performance. Given that the number of samples in other projects is small, and to prevent the impact of the reduction of samples on model training in these projects, we use 5 Fold as a reference for all projects.

Table 4: the effect of different k on the model performance

| Alg. | | Epoch | Sampling | Test percent | Acc. train | loss | Acc. | Prec. | f1 | recall |
|---|---|---|---|---|---|---|---|---|---|---|
| CNN | Infinispan | 20 | SMOTE /All | 10 | 92 | 13 | **78** | **76** | **78** | 79 |
| CNN | Infinispan | 30 | SMOTE /All | 20 | 93 | 11 | **77** | **77** | **76** | **76** |
| CNN | Infinispan | 20 | SMOTE /All | 30 | 93 | 12 | **76** | **74** | **76** | **79** |
| CNN | Infinispan | 20 | SMOTE/ All | 40 | 94 | 12 | 78 | 78 | 77 | **76** |

## 6.2 Performance Evaluation

In this section, we analyze the results of the model. Given that this is the first research in predicting the need to refer a bug to a designer, and there is no model in this area, it is not possible to compare it with previous research. Only the performance of the model can be tested using different datasets and its generality can be concluded. Precision, recall, and accuracy are used for performance evaluation. The confusion matrix is given in Table 5.

Table 6 illustrates the evaluation results. Each row of the table shows the training and test results of the model with a dataset. The results are evaluated using Accuracy, Precision, Recall, and F-Score parameters. As can be seen in the table, first of all, given that the data are balanced, the values of the different evaluation parameters for each data variance are not large. The efficiency of the algorithm varies from 55% to 78% in different datasets. In 8 out of 10 projects, precision, Recall, and F1-Score are above 75% and the variance between different parameters in each dataset is less than 2%, which is proof of the efficiency of the proposed method from data preparation to the learning model. It is noteworthy that in all projects with samples above 3000, efficiency is above 75%. In other words, if there is enough data to train the model based on deep learning, the reliable performance of the model will be proven.

Table 5: confusion matrix for the proposed classifier

|  |  | Real occurrence | |
|---|---|---|---|
|  |  | Report Need Designer | No Need for Designer |
| **Prediction** | Report Need Designer | True positive(TP) | False Positive(FP) |
|  | No Need for Designer | False Negative(FN) | True negative(TN) |

Table 6: performance evaluation of the proposed model

|  | Alg. | Project | Num. of class 1 Samples | Total sample | Class 1 Percent | Sampling | percent Test | Acc. train | Loss train | Acc. | Prec. | f1 | recall |
|---|---|---|---|---|---|---|---|---|---|---|---|---|---|
| 1 | CNN | Derby | 1101 | 4789 | 23 | SMOTE/ All | 20 | 91 | 25 | 76 | 77 | 75 | **73** |
| 2 | CNN | Infinispan | 1071 | 3977 | 27 | SMOTE /All | 20 | 93 | 11 | **77** | 77 | 76 | 76 |
| 3 | CNN | Teeid | 888 | 3797 | 23 | SMOTE/ All | 20 | 91 | 18 | **80** | 78 | 80 | 82 |
| 4 | CNN | Log4j2 | 342 | 1165 | 29 | SMOTE /All | 20 | 83 | 41 | **75** | 81 | 72 | 64 |
| 5 | CNN | Drools | 622 | 1121 | 55 | SMOTE/ All | 20 | 84 | 51 | **55** | 55 | 50 | 46 |
| 6 | CNN | HornetQ | 382 | 1039 | 37 | SMOTE/ All | 20 | 78 | 65 | **61** | 58 | 63 | 68 |
| 7 | CNN | Seam2 | 355 | 1025 | 35 | SMOTE/ All | 20 | 90 | 32 | **75** | 74 | 74 | 75 |
| 8 | CNN | Railo | 217 | 696 | 31 | SMOTE/ All | 20 | 90 | 34 | **70** | 76 | 68 | 62 |
| 9 | CNN | Izpack | 209 | 844 | 25 | SMOTE /All | 20 | 89 | 28 | **78** | 85 | 75 | 67 |
| 10 | CNN | Wildfly | 181 | 220 | 82 | SMOTE/ All | 20 | 97 | 62 | **77** | 79 | 75 | 72 |

## 7. Conclusion

In this study, the issue of bug triaging was investigated from a new perspective. Before this research, various models have been proposed to refer the processed bug to a suitable programmer based on different parameters. In this study, due to the importance of design change compared to code changes, the need to refer the bug to the designer was investigated. A CNN-based model was proposed to predict the need for a bug referral to the designer. The proposed model predicts the number of bad smells that appear in the modified software code to fix the received bug report. It uses bug report information such as summary and description, as soon as the report is received before the bug is installed and before changes are made to the software and without the need to have a modified source code in the bug fixing process. To train and test the model, 10 datasets were created using data from 10 open-source projects. The class associated with each bug report was determined using files modified in the version corresponding to that bug. The modified files in each version were checked for 16 different bad smells using the PMD tool. If a bad smell is added to the project, label "1" will be assigned to the process of fixing a bug. Label "1" means the need to refer the bug to the designer to modify the design. The model was trained and tested using these datasets. Efficiency above 75% was achieved in projects with sufficient samples and an average of 70% in all projects using this model.

## 8. Data availability

The data underlying this article will be shared upon reasonable request to the corresponding author.

**References**

[1] M. Polo, M. Piattini, F. Ruiz, Advances in software maintenance management: technologies and solutions. IGI Global, 2003.

[2] K. Sharanpreet, S. Singh, Influence of Anti-Patterns on Software Maintenance: A Review, International Journal of Computer Applications (0975 – 8887)(2015).

[3] F. Palomba, G. Bavota, M. Di Penta, et al. Detecting bad smells in source code using change history information. In Proceedings of 28th IEEE/ACM International Conference on Automated Software Engineerin (ASE'13),2013, pp. 268–278,.

[4] J. Anvik, L. Hiew, G. Murphy, Who should fix this bug?, ICSE '06 Proceedings of the 28th international conference on Software engineering, Shanghai, China, 2006.

[5] D. Čubranić, Automatic bug triage using text categorization, Proceedings o the Sixteenth International Conference on Software(SEKE), Banff, Alberta, Canada 2004.

[6] M. Rath, D. Lo, P. Mader, Analyzing Requirements and Traceability Information to Improve Bug Localization, Published in the Proceedings of the 15th IEEE/ACM Working Conference on Mining Software Repositories, (MSR) 2018, Gothenburg, Sweden.

[7] H. Xuan, T. Ferdian, L. Ming, et al., Deep Transfer Bug Localization, IEEE Transactions on Software Engineering, 2019, doi:10.1109/TSE.2019.2920771.

[8] M. Fowler, K. Beck, J. Brant, W. Opdyke, and D. Roberts, Refactoring:Improving the Design of Existing Code. Addison-Wesley,1999.

[9] F. Palomba, G. Bavota, M. Di Penta, R. Oliveto, and A. De Lucia, "Do they really smell bad? A study on developers' perception of bad code smells," in 30th IEEE International Conference on Software Maintenance and Evolution, Canada, 2014, pp. 101–110.

[10] A. Yamashita and L. Moonen, Do developers care about code smells? An exploratory survey, 2013 20th Working Conference on Reverse Engineering (WCRE), 2013, pp. 242-251, doi: 10.1109/WCRE.2013.6671299.

[11] F. Khomh, M. Di Penta, and Y.-G. Gueheneuc, An exploratory study of the impact of code smells on software changeproneness, in Proceedings of the 16th Working Conference on Reverse Engineering. Lille, France: IEEE CS Press, 2009, pp. 75–84.

[12] F. Khomh, M. Di Penta, Y.-G. Guéhéneuc, and G. Antoniol, An exploratory study of the impact of antipatterns on class changeand fault-proneness,Empirical Software Engineering, vol. 17, no. 3, pp. 243–275, 2012.

[13] M. Abbes, F. Khomh, Y.-G. Guéhéneuc, and G. Antoniol, An empirical study of the impact of two antipatterns, Blob and Spaghetti Code, on program comprehension, in 15th European Conference on Software Maintenance and Reengineering, CSMR 2011.

[14] D. I. K. Sjøberg, A. F. Yamashita, B. C. D. Anda, A. Mockus, and T. Dybåa, Quantifying the effect of code smells on maintenance effort, IEEE Trans. Software Eng., vol. 39, no. 8, pp. 1144–1156, 2013.

[15] A. F. Yamashita and L. Moonen, Do code smells reflect important maintainability aspects? in 28th IEEE International Conference on Software Maintenance, ICSM 2012, Trento, Italy, September 23-28, 2012. IEEE Computer Society, 2012, pp. 306–315.

[16] A. Yamashita and L. Moonen, "Exploring the impact of intersmell relations on software maintainability: An empirical study," in International Conference on Software Engineering (ICSE). IEEE, 2013, pp. 682–691..

[17] W. J. Brown, R. C. Malveau, W. H. Brown, and T. J. Mowbray, Anti Patterns: Refactoring Software, Architectures, and Projects in Crisis. Hoboken, NJ, USA: Wiley, 1998.

[18] P. Piotrowski, L. Madeyski, Software Defect Prediction Using Bad Code Smells: A Systematic Literature Review. In: Data-Centric Business and Applications. Lecture Notes on Data Engineering and Communications Technologies, vol 40. Springer(2020).

[19] M. D'Ambros, A. Bacchelli and M. Lanza, "On the Impact of Design Flaws on Software Defects," 2010 10th International Conference on Quality Software, Zhangjiajie, 2010, pp. 23-31.

[20] A. Lozano, M. Wermelinger, and B. Nuseibeh, Assessing the impact of bad smells using historical information, in Ninth international workshop on Principles of software evolution: in conjunction with the 6th ESEC/FSE joint meeting, ser. IWPSE '07. New Yor.

[21] D. Ratiu, S. Ducasse, T. Gîrba, R. Marinescu, Using history information to improve design flaws detection," in 8th European Conference on Software Maintenance and Reengineering (CSMR 2004), Finland, Proceeding. IEEE Computer Society, 2004, pp. 223–232.

[22] F. Palomba, G. Bavota, M. Di Penta, R. Oliveto, A. De Lucia,D. Poshyvanyk, Detecting bad smells in source code using change history information, Automated Software Engineering (ASE), 2013 IEEE/ACM 28th International Conference on, Nov 2013, pp. 268–278.

[23] M. Habayeb, S. Murtaza, A. Miranskyy, On the Use of Hidden Markov Model to Predict the Time to Fix Bugs, ieee Transactin on software engineering, 2018, 44, 12, pp. 1224-1244.

[24] R. sepahvand, R. Akbari, S. Hashemi, Predicting the Bug Fixing Time Using Word Embedding and deep LSTM, apatial issue on knowledge discovery from software repositories, IET Software, 2020.

[25] "M. Rath, P. Mäder, Influence of Structured Information in Bug Report Descriptions on IR-based Bug Localization, 2018 44th Euromicro Conference on Software Engineering and Advanced Applications (SEAA)".

[26] R. Gharibi, A. H. Rasekh, M. H. Sadreddini, S. M. Fakhrahmad, Leveraging textual properties of bug reports to localize relevant source files, Information Processing and Management 54 (2018) 1058–1076.

[27] J. Zhou, H. Zhang, and D. Lo, "Where should the bugs be fixed? more accurate information retrieval-based bug localization based on bug reports," in 34th Int. Conf. on Software Engineering, ICSE 2012, 2012.

[28] M. Tufano et al., When and Why Your Code Starts to Smell Bad (and Whether the Smells Go Away), in IEEE Transactions on Software Engineering, vol. 43, no. 11, pp. 1063-1088, 1 Nov. 2017.

[29] W. Kessentini, M. Kessentini, H. Sahraoui, S. Bechikh and A. Ouni, A Cooperative Parallel Search-Based Software Engineering Approach for Code-Smells Detection, in IEEE Transactions on Software Engineering, vol. 40, no. 9, pp. 841-861, 1 Sept. 2014..

[30] W. H. Kruskal and W. A. Wallis, Use of ranks in one-criterion variance analysis, J. Amer. Statist. Assoc., vol. 47, no. 260, pp. 583–621, 1952.

[31] Y. Xiao, J. Keung, K. E. Bennin, et al., Improving bug localization with word embedding and enhanced convolutional neural networks, Inf. Softw. Technol.,105, 2019, pp. 17-29.

[32] H. Ruan, B. Chen, X. Peng, et al., DeepLink: Recovering issue-commit links based on deep learning, Journal of Systems and Software, 2019, 158, 110406.

[33] S. Xi, Y. Yao, X. Xiao, et al. Bug Triaging Based on Tossing Sequence Modeling. J. Comput. Sci. Technol. 34, 942–956 (2019).

[34] S. R. Chidamber and C. F. Kemerer, A metrics suite for object-oriented design, IEEE Trans. Softw. Eng., vol. 20, no. 6, pp. 293–318, Jun. 1994.

[35] F. Fontana, M. Zanoni, A. Marino, M. Mantyla, Code Smell Detection: Towards a Machine Learning-based Approach. In Proceedings of the 29th International Conference on Software Maintenance (ICSM), pp. 396-399, 2013.

[36] N. V. Chawla, K. W. Bowyer, L. O. Hall, W. P. Kegelmeyer, SMOTE: Synthetic minority over–sampling technique. ournal of Artificial Intelligent Research, 16, 2002, 321- 357.

[37] E.L. Iglesias, A.S. Vieira, L. Borrajo, An HMM-based over-sampling technique to improve text classification. Expert Systems with Applications, 40 (18), 2013184-7192..

[38] H. Han, WY. Wang, BH. Mao, Borderline-SMOTE: A New Over-Sampling Method in Imbalanced Data Sets Learning. In: Huang DS., Zhang XP., Huang GB. (eds) Advances in Intelligent Computing. ICIC 2005. Lecture Notes in Computer Science, vol 3644. Springer Berlin.

[39] H. Al Majzoub, I. Elgedawy, Ö. Akaydın, et al. HCAB-SMOTE: A Hybrid Clustered Affinitive Borderline SMOTE Approach for Imbalanced Data Binary Classification. Arab J Sci Eng 45, 3205–3222 (2020). https://doi.org/10.1007/s13369-019-04336-1.

[40] Z.M. Ibrahim, M. Bader-El-Den, M. Cocea, Improving Imbalanced Students' Text Feedback Classification Using Re-sampling Based Approach. In: Advances in Computational Intelligence Systems. UKCI 2019, vol 1043, Springer, Cham.

[41] A. Gosain, S. Sardana, Farthest SMOTE: A Modified SMOTE Approach. In: Behera H., Nayak J., Naik B., Abraham A. (eds) Computational Intelligence in Data Mining. Advances in Intelligent Systems and Computing, vol 711. Springer, Singapore, 2019.

[42] X.W. Lianga, A.P. Jianga, T.Lia, Y.Y. Xuea, G.T. Wangab, LR-SMOTE — An improved unbalanced data set oversampling based on K-means and SVM, Knowledge-Based Systems, Volume 196, 21 May 2020, 105845.